\documentclass[letterpaper, 10pt, conference]{IEEEtran}
\usepackage{cite}
\usepackage{amsmath,amssymb,amsfonts}
\usepackage{algorithmic}
\usepackage{graphicx}
\usepackage{array}
\usepackage{amsmath}
\usepackage{eqnarray}
\usepackage{textcomp}
\DeclareMathOperator*{\argmax}{arg\,max}
\usepackage{xcolor}
\usepackage{comment}
\usepackage{soul}
\usepackage[subtle,tracking=normal]{savetrees}
\usepackage{subcaption}

\usepackage{xspace}
\newcommand{\sys}{SensEmo\xspace}

\def\BibTeX{{\rm B\kern-.05em{\sc i\kern-.025em b}\kern-.08em
T\kern-.1667em\lower.7ex\hbox{E}\kern-.125emX}}
    
\title{
\sys: Enabling Affective Learning through Real-time Emotion Recognition with Smartwatches}

\author{
\IEEEauthorblockN{
Kushan Choksi\IEEEauthorrefmark{1}, Hongkai Chen\IEEEauthorrefmark{2},
Karan Joshi\IEEEauthorrefmark{1},
Sukrutha Jade\IEEEauthorrefmark{1},
Shahriar Nirjon\IEEEauthorrefmark{3}, and
Shan Lin\IEEEauthorrefmark{1}}
\IEEEauthorblockA{
\IEEEauthorblockA{\IEEEauthorrefmark{1}Department of Electrical and Computer Engineering\\
Stony Brook University, Stony Brook, NY, USA\\
Email: choksi.kushan@stonybrook.edu, \{karan3502, sukrutha.jade\}@gmail.com, shan.x.lin@stonybrook.edu}
\IEEEauthorrefmark{2}Department of Information Engineering\\ The Chinese University of Hong Kong, Hong Kong SAR, China\\
Email: hkchen@ie.cuhk.edu.hk}
\IEEEauthorblockA{\IEEEauthorrefmark{3}Department of Computer Science\\ University of North Carolina, Chapel Hill, NC, USA\\
Email: nirjon@cs.unc.edu}
}

\begin{document}
\maketitle

\begin{abstract}
% Recent research has demonstrated the potential of inferring human emotions and attention responses from physiological signals measured from wearable sensors in real-time.
% Such recognition of human emotions has applications, particularly in affective learning, where it is crucial to effectively manage the emotional states of students. 
% In this paper, we introduce \sys, a smartwatch-based real-time emotion recognition system for affective learning. 
% During a class, \sys recognizes a student's levels of motivation and concentration by analyzing multiple physiological features derived from heart rate and galvanic skin responses, using a personalized emotion model that is defined by valence and arousal scales. 
% By providing real-time feedback from students, \sys employs a Markov decision process-based algorithm to enhance student learning effectiveness and experience, by offering suggestions to the teacher regarding teaching content and pacing. 
% We evaluate \sys with 22 participants in both real-world in-person and online remote classroom settings.
% The evaluation results demonstrate that \sys achieves an average accuracy of 88.9\% in recognizing student emotions.
% Furthermore, \sys assists students in attaining improved outcomes in online learning, indicated by an increase of 40.0\% in quiz grades compared to traditional learning methods that lack student emotional feedback.

Recent research has demonstrated the capability of physiological signals to infer both user emotional and attention responses. This presents an opportunity for leveraging widely available physiological sensors in smartwatches, to detect real-time emotional cues in users, such as stress and excitement. In this paper, we introduce \sys, a smartwatch-based system designed for affective learning. \sys utilizes multiple physiological sensor data, including heart rate and galvanic skin response, to recognize a student's motivation and concentration levels during class. 
This recognition is facilitated by a personalized emotion recognition model that predicts emotional states based on degrees of valence and arousal. 
With real-time emotion and attention feedback from students, we design a Markov decision process-based algorithm to enhance student learning effectiveness and experience by offering suggestions to the teacher regarding teaching content and pacing. We evaluate \sys with 22 participants in real-world classroom environments. Evaluation results show that \sys recognizes student emotion with an average of 88.9\% accuracy. More importantly, \sys assists students to achieve better online learning outcomes, e.g., an average of 40.0\% higher grades in quizzes, over the traditional learning without student emotional feedback.
\end{abstract}

\section{Introduction}

% Emotion significantly impacts the learning process, with curiosity and motivation being key factors in promoting effective learning and academic success~\cite{su2015mobile}. 
Emotion impacts learning, with curiosity and motivation being crucial for effective learning and academic success~\cite{su2015mobile}.
However, students often struggle to maintain motivation and concentration during class, especially when the difficulty level and teaching pace do not match their needs.
% While experienced teachers employ various strategies to enhance student engagement and emotional involvement, such as interactive discussions and questioning, there is a lack of effective approaches or technologies to monitor students' emotions.
Experienced teachers use strategies like interactive discussions to boost engagement, but there is a lack of effective methods to monitor students' emotions.
% This issue becomes more significant in online remote classroom settings, where teacher-student interactions are limited and thus obtaining emotional feedback from students in remote learning settings is even more challenging. 
This problem is more significant in online classrooms, where interactions are limited, making it difficult to get emotional feedback from students.
% Therefore, there is a need for an affective learning system that can accurately identify a learner's emotional state.
Thus, an affective learning system that accurately identifies a learner's emotional state is needed.
% The lack of student emotion feedback becomes an outstanding issue especially in the prevalent online teaching environments where teacher-student interactions are dramatically decreased using current remote teaching systems. 
% In fact, emotional feedback from students in remote learning environments are even harder with such few interactions. Therefore, an affective learning system that can accurately identify a learner’s emotional state is valuable.
%\hccoment{face-to-face interaction need to be connected to emotion recognition. quizzes and discussions are irrelevant. From affective learning system to previous emotion solutions.}\st{This problem becomes more severe for e-learning systems. Although e-learning systems have become essential for education under global pandemics, a recent survey shows 85\% students miss out interactions with teachers due to the limited learning environment provided by existing systems. In fact, existing e-learning systems heavily rely on quizzes and discussions after lectures to evaluate the effective learning of students, which are often delayed and incomplete.}

% Affective learning involves acquiring knowledge, skills, and attitudes through emotional engagement, acknowledging the pivotal role emotions play in cognitive processes, memory retention, and decision-making. 
Affective learning involves emotional engagement in acquiring knowledge, skills, and attitudes, recognizing emotions' role in cognitive processes, memory, and decision-making.
Research has used methods like eye tracking~\cite{barrios2004adele}, facial recognition~\cite{dhall2011emotion},
% ,busso2004analysis}
speech recognition~\cite{utane2013emotion},
% ,pan2012speech}, 
text recognition~\cite{i2020}, and gesture movement~\cite{singh2015edbl} to detect emotions.
% ,piana2014real}. 
% Despite their unique advantages, these approaches are not directly applicable or suitable for affective learning in classroom environments. 
However, these methods are not suitable for classroom settings.
For instance, camera-based methods raise privacy concerns and struggle to capture all students at once.
% camera-based methods that rely on high-resolution facial images raise privacy concerns and face challenges in capturing all students simultaneously using regular surveillance cameras. 
Additionally, in online learning, factors like lighting, network conditions, and privacy further limit these methods.
% these methods are constrained in online remote learning systems due to factors such as lighting conditions, network conditions, and privacy considerations.

%Hence our methodology using smart watch is generalized model which can be extended for multiple scenarios like Online learning, public speaking, Bipolar disorder patient-doctor analysis, etc. 

To address these challenges, we use off-the-shelf smartwatches with physiological sensing capabilities for emotion recognition. 
% Extensive studies have established strong correlations between physiological signals and user emotional and attention states
Studies have showed strong links between physiological signals and emotional states~\cite{levenson2014autonomic}. 
% The study in~\cite{new1} reviews impact and quantification of physiological sensors on user emotion. 
% It was noticed that galvanic skin resistance, skin temperature  and ECG can serve as a high correlated sensors with user emotions. 
% ~\cite{new2} contributes to the field by investigating the potential of using smartwatch signals to recognize emotions in individuals with Parkinson’s disease. 
% Leveraging these findings, we develop a personalized emotion model by learning patterns in physiological signals from different users.
% This model enables us to identify physiological signal changes related to levels of motivation and concentration.
We develop a personalized emotion model by analyzing physiological signals from different users, identifying changes related to motivation and concentration. 
% Hence, by learning physiological signal patterns obtained from users, we develop a personalized emotion model for learning physiological signal features and classifying different physiological signal changes into levels of motivation and concentration.  
Leveraging this model, we develop \sys, which helps teachers monitor students' emotions and adjust teaching methods.
% assists teachers in monitoring students' emotion states and adjusting teaching content and pace accordingly. 
For example, \sys suggests providing concrete examples to clarify key concepts when it detects confusion among multiple students. 
Similarly, it advises increasing the pace when most students appear bored. 
It is important to note that various factors can influence a student's motivation level in real-world scenarios, but this paper focuses on classroom learning situations.
% With this personalized emotion model, we develop Sensemo to assist teachers to monitor learning state of students and adapt the teaching material and pace accordingly. For example, Sensemo advises the teacher to switch to a concrete example to better illustrate a key concept when it detects multiple students are confused. Another example is that Sensemo advises the teacher to increase the pace when it detects most students are bored. We note that many factors can affect a student's level of motivation in reality, in this paper we only focus on a classroom learning scenarios.
% without interruptions.

% To assess the effectiveness of our emotion recognition solution, we implemented \sys on commercially available smartwatches.
% We conducted experiments on affective learning with 22 volunteers using the widely used International Affective Picture
% System (IAPS) emotion image dataset~\cite{lang1999international} to map physiological signal features to the valence-arousal space.
To assess our emotion recognition solution, we implemented \sys on commercial smartwatches and tested it with 22 volunteers using International Affective Picture System (IAPS) emotion image dataset~\cite{lang1999international}.
Physiological signal features collected by smartwatches were mapped to the valence-arousal space and classified into four states: curious, bored, confused, and satisfied.
% These features were then classified into four learning emotion states: curious, bored, confused, and satisfied. 
Additionally, we conducted simulated and real-world experiments to evaluate the system's impact on affective learning.
% we performed simulated and real-world evaluation experiments to assess the impact of our system on affective learning. 
% The evaluation results indicate that \sys shows promise in maintaining students' emotional state during class and enhancing their learning outcomes.
Results indicate that \sys helps maintain students' emotional states and enhances learning outcomes.
% To evaluate our emotion recognition solution, we implement Sensemo on off-the-shelf smartwatches, and conducted affective learning experiments with 22 volunteers using standard IAPS emotion image dataset~\cite{lang1999international} to map the physiological signal features to the valence-arousal space, and then classified into four learning emotion state: curious, bored, confused, and satisfied. We have also conducted simulated and real-world evaluational experiments with 22 students to test \sys in affective learning. 
% The evaluation results show that \sys demonstrates promise to sustain students' emotional state during class and improve their learning outcomes. 

In summary, our main contributions are the following.
\begin{itemize}
    \item We developed \sys, the first affective learning system that utilizes real-time emotion sensing and recognition with a smartwatch to provide students' feedback.
    \item We proposed a personalized emotion model that classifies physiological signals into specific levels of motivation and concentration tailored to the learning environment.
    % Previous research on emotion modeling has not studied emotion models in learning environments using physiological sensing data.   
    \item We introduced an affective learning model using a reinforcement learning-based controller,
    % Markov Decision Process, 
    which adapts teaching content and pace according to real-time emotion and individual learning preferences of students. This model is applicable to both online and classroom learning systems.
    % Adaptive teaching has been studied, but the incorporation of an affective model in the automatic decision making process of adaptation is novel.
    \item In real learning scenarios with 22 student volunteers, \sys achieved 88.9\% accuracy in emotion recognition. Our preliminary experiments in online and classroom settings demonstrated \sys's promise to boost learning outcomes and enhance student emotions.
\end{itemize}

\section{Related Work}

% Emotion recognition has been extensively explored in various domains. 
% A KNN algorithm is used to classify emotions using both physiological and subjective components associated with emotions
Emotion recognition has been explored in many domains. A KNN algorithm classifies emotions using physiological and subjective components~\cite{nasoz2004emotion}. 
StressSense uses smartphone-detected voice for stress recognition~\cite{lu2012stresssense}. 
Text analysis studies emotions via semantic labels, attributes~\cite{wu2006emotion}, and typing rhythms~\cite{epp2011identifying}.
% Emotion recognition in text analysis has studied emotional states represented by sequences of semantic labels and attributes~\cite{wu2006emotion}, as well as by typing rhythms~\cite{epp2011identifying}. 
% Recent advancements in wearable devices have enabled the use of physiological signals, such as respiratory frequency, heart rate variability, skin conductance, and accelerometer data, to detect human emotions
Wearable devices detect emotions using respiratory frequency, heart rate variability, skin conductance, and accelerometer data~\cite{hernando2016inclusion}. 
However, limited research focuses on emotion recognition in learning environments. Studies explore wearable biosensors to enhance learning by monitoring physiological signals~\cite{new3}, assessing cognitive load during problem-solving~\cite{romine2020using}, and predicting depression in students~\cite{wang2018tracking}.
% However, there has been limited research on emotion sensing and recognition specifically in learning environments.
% Studies examines the potential of wearable biosensors, including electrodermal activity, heart rate variability, and electromyography, to enhance learning environments by monitoring physiological signals~\cite{new3}.
% Some studies leverage wearable devices to assess cognitive load during problem-solving activities~\cite{romine2020using} or predict depression in college students~\cite{wang2018tracking}. 
However, few have examined emotion recognition's role in improving learning performance.
% limited work has explored the application of emotion recognition in improving learning performance.

% Through meticulous analysis, the review sheds light on the potential of wearable biosensors to enhance learning environments by monitoring physiological signals, thus offering valuable insights into the future of educational technology. In Sensemo, we use physiological signal from smartphones and machine learning methods to build a mapping from data to valence-arousal space. In addition, existing work did not study any affective learning systems that use the emotion recognition to improve learning performance.

Adaptive systems have been proposed to adjust game difficulty based on users' emotions to maintain engagement~\cite{chanel2011emotion}, and e-learning systems have used neuro-fuzzy networks to estimate user behavior and adapt content presentation~\cite{asteriadis2009estimation}. 
% In other domains, adaptive systems have been proposed to adjust game difficulty based on users' emotions to maintain engagement~\cite{chanel2011emotion}, and e-learning systems have used neuro-fuzzy networks to estimate user behavior and adapt content presentation~\cite{asteriadis2009estimation}. 
% A recent study introduces an approach for extrinsic emotion regulation using an affective intelligent agent model~\cite{pico2024towards}. 
However, they are usually limited by real-world implementation and validation. 
In contrast, \sys utilizes physiological signals from smartwatches 
% and machine learning techniques to map data to the valence-arousal space. 
% It also employs an adaptive online learning system based on Markov Decision Process 
to guide teaching content and pace 
in real-world settings.
% according to the emotions of a group of students, aiming to improve overall student learning performance.

\section{System Design}\label{sec:sys}

\subsection{Application Scenarios}
\sys can be implemented in both in-person and online remote classroom settings,
where students wear a smartwatch, and a mobile app collects real-time physiological data.
This data is then sent to a central server for emotion recognition, adapting the learning process for students.
% Based on the real-time emotion recognition results, \sys adapts the learning process for students. 
In online remote learning, teaching content and pace can be automatically adjusted for each individual student.
In an in-person classroom setting, the instructor focuses on maintaining positive emotional responses among students.
% The learning system adaption functions are different in these two scenarios: for online learning, the system monitors an individual student's emotion response and conduct adaptation, i.e. changing speed or content, either automatically or manually by the teacher. A teacher can monitor the status of system and override the actions of the system if needed. This requires specially designed teaching material and prerecorded segments of lectures. For classroom settings, a lecturer can perform various interactive learning based on students' emotion responses. The lecturer needs to sustain the positive emotional response of the majority of students, also pay attention to individuals who are less positive. 
% This paper primarily explores the use of \sys in affective learning, but it also has the potential to be applied in other scenarios, including public speaking, interviewing, and smart TV.
This paper explores \sys's use in affective learning, with potential applications in driving monitoring~\cite{huang2019magtrack}, health monitoring~\cite{li2024emomarker}, and interviewing/consultation assistance~\cite{yang2024drhouse}.

\subsection{System Overview}

\begin{figure}[t]
\centerline{\includegraphics[width=0.8\linewidth]{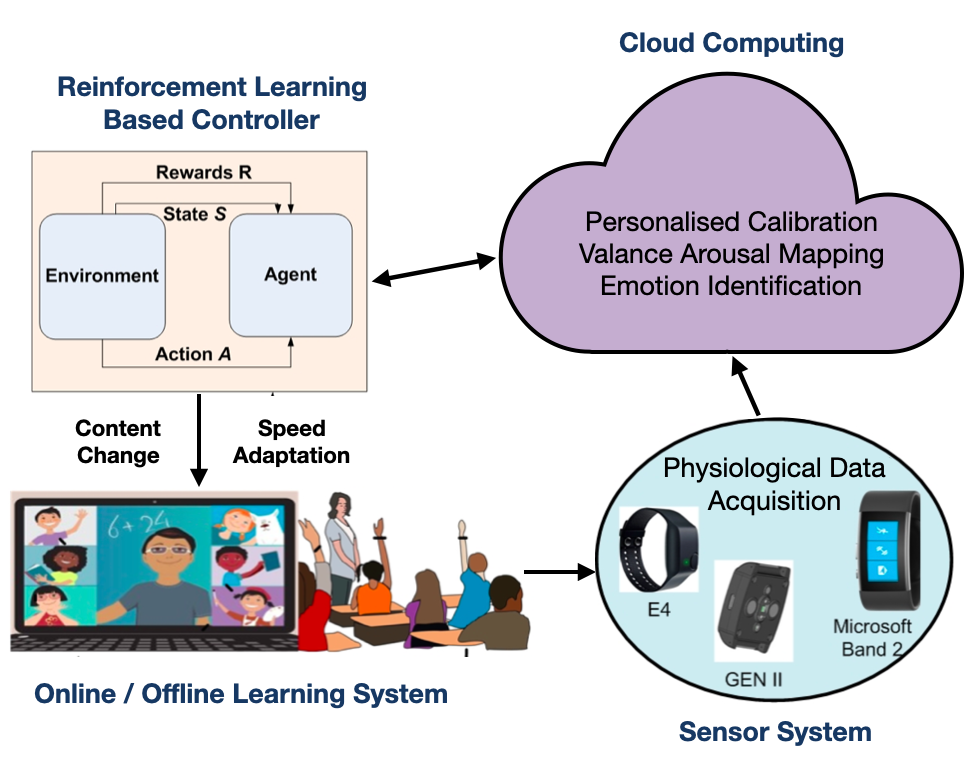}}
\vspace{-1ex}
\caption{
% System overview of \sys. It utilizes physiological signals to learn a personalized emotion model and an adaptive learning system to guide teaching content and pace according to student emotion.
\sys uses physiological signals to personalize the emotion model and adapt the learning content and pace to the student's emotional state.}
\vspace{-0.1in}
\label{system}
\end{figure}

% \sys utilizes real-time user feedback by collecting emotions from the student's physiological signals
\sys collects real-time emotional feedback from students' physiological signals during engagement with teaching content. This feedback is used to adapt the course automatically. The components of \sys are shown in Figure~\ref{system}.
% while they are actively engaged with teaching content.
% This emotional feedback is then used to automatically adapt the course. 
% The components of \sys, illustrated in Figure~\ref{system}, are the following.
% The key components of the system as depicted in Figure \ref{system} ar

% \begin{itemize}
% \item  
% \vspace{0.05in}
\noindent \textbf{Sensor system}. 
\sys's sensor system 
% involves utilizing a smartwatch 
uses a smartwatch to capture real-time physiological signals, including heart rate, skin resistance, and skin temperature.
% (e.g., TAG Heuer Connected Calibr E4, Microsoft Band 2) 
% to capture real-time physiological signals from the user, including parameters like heart rate, skin resistance, and skin temperature. 
% a smartwatch is used to record a user’s real time physiological signal, such as heart rate, skin resistance, skin temperature etc. 
% Many smartwatches such as E4, GEN II, and Microsoft Band 2 have physiological sensors to perform this desired sensing function. 

% \vspace{0.05in}
\noindent \textbf{Physiological data collection}. 
A smartphone application communicates with smartwatches to retrieve sensor data, which is sent to the cloud for processing and feature extraction.
% A smartphone application establishes communication with smartwatches and retrieves sensor data from them. 
% The collected data is then sent to the cloud for further processing, where various features are extracted.
% as described in the subsequent subsection. 
% to collect data from the band, we build an android app which communicates with the smartwatch and collect sensor data. Moreover, this app processes the data for various feature extractions as discussed in the following subsection. The app then transfers the data to the cloud.

% \vspace{0.05in}
\noindent \textbf{Cloud computing}. 
A cloud platform is used due to storage and computational requirements, facilitating feature mapping, emotion recognition, and controller computations.
% cloud storage is used to manage large data in terms of storage and computation. Feature mapping, emotion recognition and controller computations are mainly accomplished on cloud platform.

\begin{figure}[t]
\centerline{\includegraphics[width=0.9\linewidth]{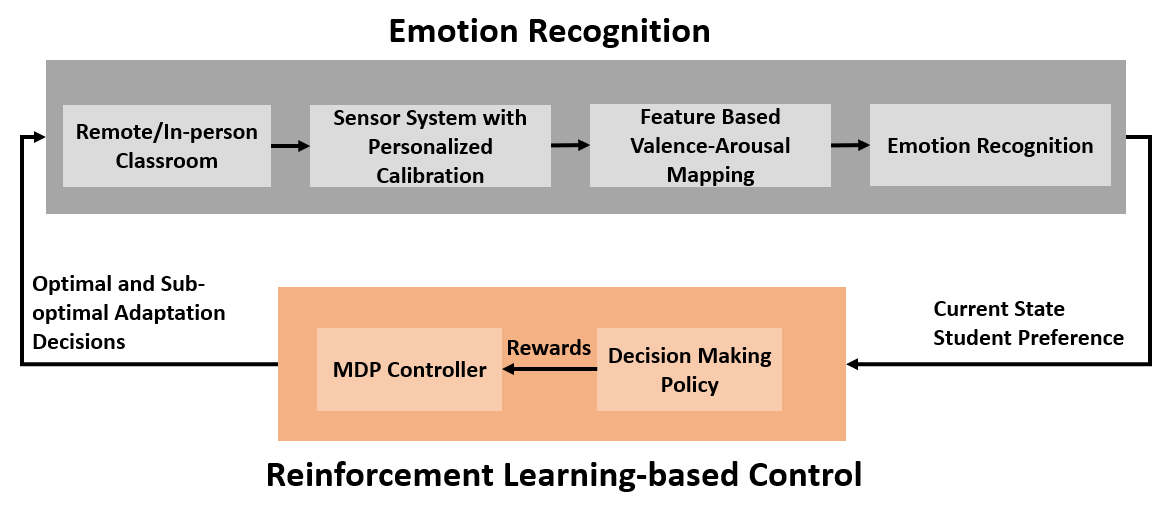}}
\vspace{-0.05in}
\caption{Overview of the affective learning feedback controller, including emotion recognition and a reinforcement learning-based controller.}
\vspace{-0.2in}
\label{fbc}
\end{figure}

% \vspace{0.05in}
\noindent \textbf{Affective learning controller}. 
% The learning controller utilizes real-time feedback on emotions and concentration to make decisions regarding 
The learning controller uses real-time feedback on emotions and concentration to \textit{adjust teaching content and pace}.
% \textit{adjustments in teaching content and pace}. 
It employs a Markov Decision Process (MDP) trained via reinforcement learning, 
creating a loop that informs the instructor of students' emotions and preferences.
% establishing a closed loop that provides feedback to the instructor based on the current emotions and preferences of the students.
% The learning controller uses the real-time emotion and concentration feedback to make decision for the teaching content and pace.
% Markov Decision Process (MDP) is trained based on reinforcement learning, completing the closed loop and providing feedback to lecturer about \textit{adaptation in pace and content} based on student current emotion and preferences. 
% \end{itemize}
The design, shown in Figure~\ref{fbc}, includes emotion recognition and reinforcement learning-based control.
% Figure~\ref{fbc} illustrates the overview design of the affective learning controller, which consist of two major components (i.e., emotion recognition and reinforcement learning-based control) that we will discussed later. 

\subsection{Emotion Recognition}

During the students' engagement in learning, \sys 
% utilizes smartwatches to capture physiological data. 
uses smartwatches to capture physiological data. 
To interpret this data, it is crucial to correlate physiological signals with emotional states. 
In this section, we describe our approach to emotion recognition.
% To correctly interpret this data, it is crucial to identify and establish a correlation between the various physiological signals and emotional states. 
% In this section, we will describe our approach for emotion recognition.

% \subsubsection{Physiological feature selection}
% \vspace{0.05in}
\noindent \textbf{Physiological feature selection}.
Emotional responses trigger autonomic nervous system reactions, captured through physiological signals.
% Emotional responses are consistently accompanied by autonomic nervous system reactions, which can be captured by employing appropriate features of physiological signals. 
\sys uses the following features to deduce emotions. (1)~Electrodermal activity: Continuous changes in skin's electrical characteristics, influenced by moisture.
\textit{Skin conductance response}, increasing with stress and perspiration, and
% It is measured as the \textit{skin conductivity response}, which increases when individuals experience stress and perspiration. 
\textit{skin conductance level} can be obtained from electrodermal activity.
(2)~Blood volume pulse: Captures changes in blood volume, indicating blood flow. 
The inter-beat interval, which represents the time duration between two heartbeats, is derived from the blood volume pulse signal. 
This interval is utilized to determine features such as \textit{heart rate} and \textit{heart rate variability}. %Heart rate and heart rate variability are  reliable index of emotional states \cite{appelhans2006heart}. 
(3)~Skin temperature: Linked to the autonomic nervous system.
Excitement can raise skin temperature. 
We measure both \textit{skin temperature response and level}.
% When a person becomes excited, their skin temperature tends to rise. Our system measures both the \textit{skin temperature response} and \textit{skin temperature level}. 
 % \end{itemize}
% These signals are obtained from the built-in sensors on smartwatch and then many other features are extracted further to be used into our study. We explain about these signals and features in detail in the system implementation section.
% All experiments conducted in this paper received an IRB waiver for ethical approval, which is allowed according to the Standard Operating Procedures at authors' affiliated universities.

% \subsubsection{Personalized emotion model calibration}
% \vspace{0.05in}
\noindent \textbf{Personalized emotion model calibration}.
% One of the significant challenges in emotion recognition using physiological signals is the variation in physiological responses among individuals experiencing the same emotion
Emotion recognition using physiological signals varies  among individuals~\cite{gottman1995relationship}. 
%There are two main types of variations. The first type relates to baseline differences, such as different individuals having regular heart rates of 65 bps and 75 bps, respectively, which is unrelated to emotion. The second type is related to physiological response to specific emotions, where 
For instance, one person's heart rate may rise to 100 bps when excited, another's to 120 bps.
% For example, one person's heart rate may rise to 100 bps when excited, while another person's heart rate may easily reach 120 bps under the same emotion state. 
To address the user-specific differences, we developed a personalized calibration technique to normalize emotion-reflecting physiological changes.
% we develop a personalized calibration technique to normalize the extent of emotion changes reflected in the user's physiological signal variations.
% So we use a normalization technique to map the physiological signals to a feature space [-1,1].
Our local min-max normalization maps physiological signals to a feature space from -1 to 1, calculated as
% Specifically, we propose a local min-max normalization approach that maps physiological signals to a feature space ranging from -1 to 1.
% It improves efficiency of emotion recognition as well as helps in tracing the user response to \sys for better training adaptive feedback learning. 
% The normalized feature $f'$ is computed as
% \begin{equation*}
    % f'=\frac{f-{f_{min}}}{{f_{max}}-{f_{min}}}
% \end{equation*}
    $f'=(f-{f_{min}})/({f_{max}}-{f_{min}})$,
where $f$ is the feature, ${f_{max}}$ and ${f_{min}}$ are local maxima and minima.
This normalization has been used to analyze different physiological features consistently, disregarding individual emotional responses~\cite{henderi2021comparison}. 
% For instance, STR of user 1 is having a huge value compared to user 2. However, it can be observed after normalization that both user 1 and user 2 wnt through emotion attribute variation based on their own personal threshold. We can compare this with averaging approaches, where emotional attribute change of user 1 will dominate the decision of \sys.

% It shall be noted that user feature may vary drastically if $X$ is used directly owing to nature of physiological feature, whereas it become bounded between -1 to +1 standardizing physiological feature. 

% \subsubsection{Valence/arousal scales from physiological features}
% \vspace{0.05in}
\noindent \textbf{Valence/arousal scales from physiological features}.
% Once, personalized thresholds of features are available, one needs to map this features to emotional states in standard validation manner. 
\sys predicts students' valence and arousal scales from calibrated features.
We use the IAPS image dataset, which elicits emotional responses.
% We utilize the IAPS image set, which contains images meant to elicit an emotional response from the user.
Images are categorized as pleasant, neutral, or unpleasant, and rated on valence (pleasant to unpleasant) and arousal (calm to excited) scales.
% Images are categorized as pleasant, neutral or unpleasant, and each image is assigned valence (ranging from pleasant to unpleasant) and arousal (ranging from calm to excited) scales rated by researchers. 
% These scales map physiological features into the valence-arousal space, widely used in psychological research.
These scales are target values for mapping physiological features into the valence-arousal space, widely used in psychological research~\cite{choi2017heart}. 
% During our study, users' physiological signals are collected while they are exposed to these images. 
During our study, users' physiological signals were collected while viewing these images. 
This collected dataset is used to train a model to translate physiological signals into valence and arousal scales.
% The collected physiological signals, along with their corresponding valence and arousal scales, are used as a dataset to train a machine learning model capable of translating physiological signals into valence and arousal scales.
% We used SVM to train the model.  
% The model is validated with test set images. The evaluation results are shown as follows. 

% This photo-set has pictures rated for the valence and arousal values. These data points, recorded features and the rated values of arousal and valence, make the learning set for sensor data tuning using supervised learning algorithm. 
% Hence, such a methodology allows us to tune the sensor system data to correct valence-arousal thresholds ensuring optimal mapping of emotional states into valence-arousal space for the participants. 

\begin{figure}[t]
\centerline{\includegraphics[width= 0.75\linewidth]{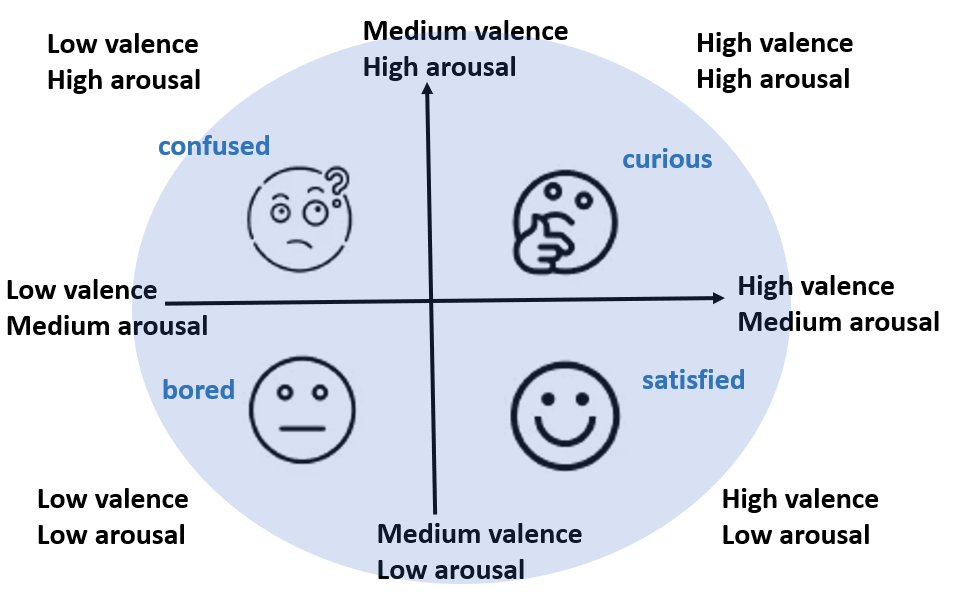}}
\vspace{-0.0in}
\caption{The 2-dimensional valence-arousal space and mapping of emotions of bored, satisfied, curious, and confused.}
\vspace{-0.1in}
\label{vaj}
\end{figure}

%\revise{
% \subsubsection{Emotion recognition from valence-arousal space}
% \vspace{0.05in}
\noindent \textbf{Learning emotion recognition from valence-arousal space}.
The valence-arousal space categorizes emotions in a two-dimensional space~\cite{lang1995emotion}. 
Existing theories label emotions within this space~\cite{posner2005circumplex}. 
% To capture emotions, \sys employs a mapping from valence and arousal scales to learning-related emotions such as boredom, confusion, curiosity, and satisfaction
\sys maps valence and arousal scales to learning-related emotions like boredom, confusion, curiosity, and satisfaction~\cite{kort2001affective}. 
Figure~\ref{vaj} shows the valence-arousal space highlighting learning-relevant emotions.
% depicts a diagram of the valence-arousal space highlighting the emotions relevant to learning. 
% This approach, which allows for the continuous modeling of emotions, has been validated using clinical experimental data
This approach, allowing continuous emotion modeling, has been validated by clinical data~\cite{mandryk2007fuzzy}.
 % We use a grid to map physiological data onto three levels of arousal and valence. 
% This approach for continuously modeling emotions has been validated using real experimental data~\cite{mandryk2007fuzzy}.

% Our methodology refer to the Russel \cite{posner2005circumplex}, and Ekman and Freisen's \cite{ekman1987universals} theory of emotions which provide discrete tags to emotions, as depicted in the Figure \ref{vaj}, for building our system model. We also refer to Kort's learning spiral model \cite{kort2001affective} which proposes emotions that are related to learning, such as boredom, confusion, curiosity and satisfaction. Our system maps the derived features from the sensor data onto the valence-arousal space. We divide our arousal and valence space into three levels each and these levels are mapped to discrete emotions based on study conducted by \cite{mandryk2007fuzzy}, which provides fuzzy rules for mapping each level to particular emotions. Our model transforms these fuzzy rules into discrete rules. Figure \ref{vaj} shows our emotion mapping model and depicts the rules used to map arousal-valence levels to emotions.

\subsection{Reinforcement Learning-based Control}
\begin{figure}[t]
    \centering
    \centerline{\includegraphics[width= .8\linewidth]{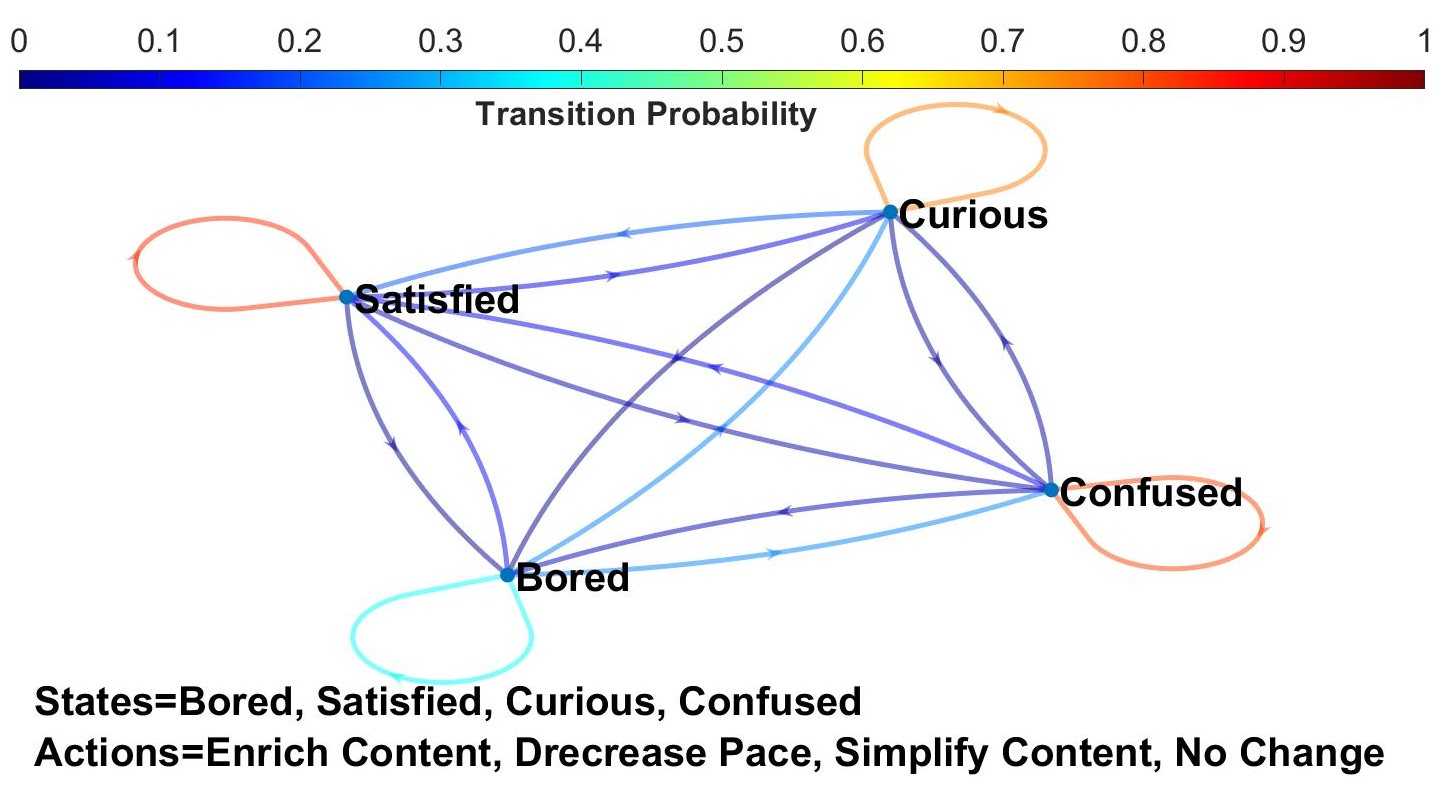}}
     \vspace{-0.0in}
    \caption{The discrete MDP representation of \sys.}
     \vspace{-0.2in}
    \label{MDP structure}
\end{figure}

% While involved in learning activities, a student can go through many different emotions, but these can be broadly classified into four major learning related emotions - satisfied, boredom, curiosity and confusion. 
% The feedback control model can be logically divided into two subsections- Emotion sensing system and feedback control decision maker, which has been depicted in Figure \ref{fbc}.
% After the emotion recognition system processes student emotion data, the resulting output includes the current emotional state, denoted as
After processing student emotion data, the system identifies the current emotional state, $S_t$, and preferences for content difficulty and pace.
% This emotional state and the student's preference for specific choices related to content difficulty (such as visual or text) and pace. 
% These inputs are then provided to a reinforcement learning-based controller.
% After student emotion data is processed by emotion recognition system, the current emotion state $S_t$, as well as student preference, is given as input to a reinforcement learning-based controller.
% Student preference is defined as user specific choices regarding change of content difficulty (e.g., visual or text) and pace. 
% The reinforcement learning-based controller is composed of two main components: a decision-making policy lookup and a MDP controller. 
These inputs are fed to a reinforcement learning-based controller with two main components: a decision-making policy lookup and an MDP controller.
The lookup retrieves information about optimal and sub-optimal decisions made by the MDP controller in the current and previous steps.
% The decision-making policy lookup retrieves information about both the current and previous optimal as well as sub-optimal decisions made by the MDP controller.
Sub-optimal decisions are fallback options if the optimal policy isn't feasible.
% Sub-optimal decisions refer to secondary decisions that the system can take if the optimal policy is not feasible in a practical scenario. 
For instance, lectures may not able to change content or adjust pace as suggested by the MDP controller.
The MDP controller optimizes decision sequences based on the current emotional state $S_t$ and student preferences, aiming to keep the user in a state of curiosity, characterized by relaxed alertness~\cite{caine201512}.
% The MDP controller optimizes its iterations of the Markov model to determine the best course of action, considering both the current emotional state $S_t$ and the student's preference. 
% This optimization process produces both optimal and sub-optimal decisions that guide the system's actions.
% The mathematics and operation of MDP controller are given in greater detail in later section. 
% The objective of \sys is to maintain the user in an \textit{optimum emotion state of curiosity}, which is characterized as a state of relaxed alertness~\cite{caine201512}, thereby promoting optimal learning. 
% This is achieved by adapting the teaching content and pace based on emotional feedback.
The MDP structure is shown in Figure~\ref{MDP structure}, comprising states (bored, satisfied, curious, confused) and actions (increase/decrease pace, simplify/no change in content). 
% In this MDP, student emotional states are defined as a set of emotions, specifically: bored, satisfied, curious, and confused. 
% The decisions available to the instructors include increasing pace, decreasing pace, simplifying content, or making no change.
% The transition probability determines the likelihood of moving from one emotional state to another. 
Transition probabilities determine the likelihood of moving from one state to another, influenced by the current state and the student’s preferences.
% Decisions are made by considering both the transition probability and the student's current emotional state.
However, determining the optimal decision that maintains the user in the desired emotional state of curiosity relies on value iteration of the MDP model.

% After the emotion recognition through the sensor data, we now have access to the student's emotional state which can help us make our system adapt so that we can promote optimum learning.
%The feedback control system, depicted in Figure \ref{fbc}, can be considered as a black box which takes as input the current inferred emotions and provides as output the decision to change speed or content of the slides. These decisions are based on the current emotion state and the desired emotion state, curiosity. For example, if a student is in boredom state so we can change the content of the slide to introduce more illustration based content for same topic, assuming that we have this illustrative content. Similarly, we can make decisions on speed of slides. This decision making implementation is based on a Markov decision process.

% \vspace{0.05in}
% \noindent\textbf{Markov decision process.}
\sys faces uncertainties in both the students' emotional states and learning performance, which are influenced by the instructor's decisions.  
The system uses a MDP represented by a 4-tuple: states $S$, actions $A$, transition probabilities $P$, and rewards $R$.
% To address this, we define an optimization problem for the MDP, aiming to find an optimal policy denoted as $\pi^*$ for the controller's decision-making process. 
% This policy serves as a function that determines the appropriate decision to be taken in order to evoke an emotional state of curiosity in students, considering their current emotional state. 
The goal is to find an optimal policy $\pi^*$ that maximizes cumulative rewards over time, guiding decisions to induce curiosity.
% In particular, our objective is to select a policy that maximizes the cumulative rewards over a potentially infinite or asymptotic time horizon.
% This complexity necessitates the selection of a MDP controller, as it is well-suited to handle dynamic learning environments and continuous feedback, enabling optimal decision-making.
%Markov Decision Processes is a discrete stochastic decision making extension of Markov chains used when outcomes are partly random. MDP must follow Markov property which compels the stochastic decision making process to be memory less and dynamic which help the feedback controller to be near real time controller \cite{chen2011adaptive}.
% We formulate an optimization problem for MDP to find a optimal policy $\pi^*$ for controller decision making: a function that would specify the action to be taken to achieve curious emotional state for students given student current emotional state. 
% In other words, our objective is to choose a policy that will maximize cumulative function of the rewards over a potentially infinite/asymptotic horizon. 
% The MDP representation of the \sys system is depicted in Figure~\ref{MDP structure}.
% It consists of a 4-tuple comprising the following elements: the set of states $S$, the set of actions $A$, the transition probabilities $P_a$, and the associated rewards $R_a$.
% These four states are derived using emotion recognition which can be given as Bored, satisfied, Curious and confused. 
% The MDP structure can be formally defined as follows.
The MDP is defined as follows.

\begin{itemize}
    \item \textbf{States}: $S = \{s_0, s_1, s_2, s_3\}$.
  % (bored, satisfied, curious, confused).
    % The emotional state space is composed of four distinct states, denoted as $S =\{s_0, s_1, s_2, s_3\}$. 
    Each represents emotion state \textit{bored, satisfied, curious, and confused}, respectively. 
    Additionally, $s'$ denotes the state that follows after $s$.
    % Also, $s'$ represents the subsequent state after $s$.
    % The emotional state space $S$ = \{$s_0$,$s_1$,$s_2$,$s_3$\}. Such that $s_0$, $s_1$, $s_2$ and $s_3$ represents bored, satisfied, Curious and confused state respectively. $s$ and $s'$ suggest current and next state in derivation.
    % It shall be noted that we have divided student emotional states into four categories as seen in Figure \ref{MDP structure} and are represented as set space $S$ = \{$s_0$,$s_1$,$s_2$,$s_3$\}.
    \item \textbf{Actions}: $A = \{a_0, a_1, a_2, a_3\}$.
  % (increase pace, decrease pace, change content, no change).
    % The action state space $A= \{a_0,a_1,a_2,a_3\}$ consists of four actions: 
    Each represents action \textit{increase pace, decrease pace, simplify content, and no change in content}. 
    Note that $a$ is the action taken during each iteration, as indicated in equations \eqref{policy} and \eqref{valueiter}.
    % Action state space $A= \{a_0,a_1,a_2,a_3\}$ are increase/decrease pace, and change/no change content. It shall be noted that $a$ represents immediate action for each iteration as suggested in \eqref{policy}-\eqref{valueiter}.
    % Similarly, set of action includes increasing speed (Up), decreasing speed (Down), Simplify Content and no change are mathematically represented as $A$= \{$a_0$,$a_1$,$a_2$,$a_3$\}.
    \item \textbf{Transition Probabilities}:
    % The probability of transitioning from state $s$ to $s'$, given a specific action $a$, is d
    Determined by a Markov chain asymptotic analysis~\cite{seneta2006non} with our experimental data.
    % Transition probability is probability of the Markov process changing from one state to another state. 
    % $P_a (s_{t+1}|s_{t},a)$ is probability of transition from state $s_t$ to $s_{t+1}$ given action $a$ is computed using Markov chain asymptotic analysis on our experimental setup data as explained in later section.
    % The Markov chain asymptotic analysis~\cite{seneta2006non} is utilized to determine the characteristics of the designed Markov chain. 
    % We assume the probability of state transitions to be the asymptotic state transition probability.
    % determines the properties of designed Markov chain. In this approach we have assumed our probability of state transition as the asymptomatic state transition probability.
    \item \textbf{Reward}: $R_{a}(s|s')$. 
    It represents the incentives for state transition. 
    The initial rewards were defined based on the observed differences in asymptotic state transition probabilities.
    % , as depicted in Figure~\ref{MDP structure}.
    % For instance, the reward $R_{a_{0}}(s_0|s_1)$ was set to 0.4, $R_{a_{0}}(s_0|s_2)$ was set to 0.25, and $R_{a_{0}}(s_0|s_3)$ was set to 0.1. 
    These values were obtained empirically, considering the variations in state transition probabilities.
    % obtained during the initial validation of the data.
    % Rewards $R_{a}(s|s')$ represents expected immediate incentives associated with the algorithm for stabilizing on the next stage $s'$. The initial rewards were defined in proportion of asymptomatic state transition probability differences as observed from Fig. \ref{MDP structure}. Such that  $R_{a_{o}}(s_o|s_1)$ =0.4, $R_{a_{o}}(s_o|s_2)$ =0.25, and $R_{a_{o}}(s_o|s_3)$ =0.1. The look up table was established empirically, taking into account the variances in state transition probabilities derived from the initial validation of the data truth.  
\end{itemize}

%In this work, we deal with partially observable MDP as we have knowledge of students current state $S_t$ based on emotion identification. 

The optimization problem for the MDP aims at maximizing the accumulated rewards based on current emotional states of the students, in order to determine the optimal policy $\pi^*$.
The optimal policy $\pi^*(s)$ is determined asymptotically using the value iteration optimization function, which is defined by the following equations.
\begin{align}
 \pi(s) &:=  \argmax_a\{\sum_{s'}P(s'|s,a)(R(s'|s,a)+\gamma V(s'))\}
\label{policy}\\
 V_{i+1}(s) &:=  \max_{a}\{\sum_{s'}P_{a}(s'|s)(R_{a}(s,s')+\gamma V_{i}(s'))\}
\label{valueiter}
\end{align}

The policy $\pi(s)$ maps states $s$ to actions $a$ and 
the value function $V_{i+1}(s)$ at iteration $i+1$ represents the maximum expected return from state $s$, subject to the optimal action and the initialization of the value function. 
% during value iteration employs the Bellman equation~\cite{o2018uncertainty}. 
% It represents the maximum expected return achievable from state $s$.
% The value function $V(s')$ corresponds to the maximum expected value of state $s'$ based on the optimal action and the initialization of the value function. 
Initial values for states are assigned randomly. 
% For the states $s_0$, $s_1$, $s_2$, and $s_3$, the value function $V(s')$ is initially assigned random values.
The discount factor $\gamma \in [0,1]$ weighs the importance of future rewards, empirically set to guide actions towards curious or satisfied states. 
% Empirically, the discount factors were selected to guide actions towards achieving curious or satisfied states. 
In our system, the discount factors for $s_1,\ldots,s_4$ are set to $0.1$, $0.45$, $0.35$, and $0.1$, respectively.
Furthermore, the decision-making process takes into account the student's preferences, such as desirable learning pace or preference between illustrations and descriptions. 
This preference data is collected through a pre-lecture questionnaire that can be conveniently stored online.
% Here, $\pi(s)$ represents the policy function that maps states $s$ to action $a$.
% The value function $V_{i+1}(s)$ of state $s$ at iteration $i+1$ during value iteration uses Bellman optimality equation. 
% The maximum value function, representing the expected maximum return achievable from state $s$. 
% $V(s')$ is maximum expected value of $s'$ based on optimal action and initialization of value function. 
% The initialization of value function $V(s')$ for states $\{s_0,s_1,s_2,s_3\}=\{0,0.5,0,0\}$ respectively. 
% This values were chosen randomly. 
% The value function $V(s')$ for states $\{s_0,s_1,s_2,s_3\}$ is initialized randomly.
% Lastly, $\gamma$ is the discount factor which takes value from 0 to 1. 
% The discount factors were empirically selected such that actions would lead to curious or satisfied states. 
% Hence, the discount factor for bored, satisfied, curious and confused states were 0.1, 0.45, 0.35 and 0.1 respectively. The decision are at the same time also based on the student's preferences, like if he/she is a slow learner or prefers illustrations over descriptions. This student preference data is collected by a questionnaire before the start of the lecture, which can be easily stored online. More detailed explanation can be seen in the following section.

\section{System Implementation}
% System implementation gives a detailed description of how physical design and information flow was executed for Sensemo system.
In this section, we describe the implementation and realization of both the hardware and algorithms of \sys.
% physical design and information flow in sensor system, emotion recognition and control components of the \sys system.

\subsection{Sensor System}
% The sensor system implementation mainly deals with data management and generation for any further data segmentation and decision making process. It can be divided into data acquisition and feature extraction.
% We implement the sensor system of \sys with off-the-shelf smartwatches.

\noindent \textbf{Data Acquisition}.
We implemented \sys on Microsoft Band 2 smartwatches, which can detect electrodermal activity, blood volume pulse, and skin temperature.
Each sensor has a default sampling rate of $1$ Hz. 
Proper sensor placement is crucial to avoid motion artifacts that can affect measurement accuracy.
% Each sensor has its own sampling rate, with the default sampling rate set to 1 Hz.
% It is important to consider sensor placement to ensure reliable data collection, as motion artifacts can affect the accuracy of the measurements.
An Android app was installed on users’ smartphones to receive sensor data from the smartwatches and facilitate data collection.
% Furthermore, an Android application was installed on users' smartphones. This application is responsible for receiving the sensor data transmitted from the smartwatches and facilitating the data collection process.

% \vspace{0.05in}
\noindent \textbf{Feature extraction}. \sys extracts six features from the sensor data: skin conductance response (SCR), skin conductance level (SCL), heart rate (HR), heart rate variability (HRV), skin temperature response (STR), and skin temperature level (STL). 
It computes moving averages of HR, STR, and SCL, and running deviations of HRV, SCR, and STR using a 50-sample moving window. 
% the running deviation of HRV, SCR, and STR using a moving window of length 50 samples, which represents the local standard deviation over time. 
HRV is extracted from the RR-interval data stream using the Root-mean-square of successive differences (RMSSD) method.
% HRV is specifically extracted from the RR-interval data stream using the Root-mean-square of successive differences (RMSSD) method.

% \sys extracts \textit{six} different features, namely skin conductance response (SCR), skin conductance level (SCL), heart rate (HR), heart rate variability (HRV), skin temperature response (STR), and skin temperature level (STL). Sensemo computes moving averages of HR, STR and SCL, and running deviation (the local standard deviation using a moving window over time) of HRV, SCR and STR, in a window of length 50 samples. 
% HRV is extracted from the RR-interval data stream using the Root-mean-square of successive differences (RMSSD) method.

% The android app, after collecting the sensor data and computing all the features, then stores the data to a cloud database which then can be easily accessed by the e-learning application.

\subsection{Emotion Recognition}
% Having recognized the features and signals to be used for the system, it is essential to devise a technique to infer emotion from these signals and features. There has been limited research in field of emotion recognition using physiological signal, unlike other methods such as face recognition, and thus there is limited knowledge about what levels or values of physiological signal accompany which emotion. 
% However, there is one major challenges posed during calibration of physiological signal as discussed in system design for sensor system tuning.
We describe two major components of emotion recognition: personalized model calibration and emotion recognition.

% \vspace{0.05in}
\noindent \textbf{Personalized model calibration}.
%   In other words, As we know physiological signal were collected from the smart watch; however data did not allow us to infer the emotions on the valence arousal space. 
% It is observed that physiological sensor data baselines and thresholds of different users are different, so personalized calibration of data is necessary.
% This leads us find a solution which is generalized in order to label all the user emotion into valance arousal space.
First, \sys calculates the maximum and minimum values of the six physiological sensor features (i.e., SCR, SCL, HR, HRV, STR, and STL) within a 50-sample window for each user.
Then, it computes the normalized features for all users. 

% \vspace{0.05in}
\noindent \textbf{Emotion recognition}.
%  Emotion identification deals inference of emotion from the sensor data and extracted features using machine learning. Earnest common knowledge demands emotions to be segmented as optimally even in noisy environment and should be relatively memory efficient.
%  This leads us to use fine Gaussian SVM with and without personalized calibration. 
% We leverage the IAPS image dataset to map the normalized features to valence and arousal scales.
We use the IAPS image dataset to map normalized features to valence and arousal scales.
During data collection, users' physiological signals were recorded while viewing images, with the valence and arousal scales of those images considered as the ground truth.
% , and the valence and arousal scales associated with those images were considered as the ground truth of the users' valence and arousal scales.
% emotions in the valence arousal space. 
% Lastly, \sys labels the dataset using IAPS data thresholds. 
% Our solution could also generalize to various users with high accuracy.
% We use machine learning approaches to infer emotions from the sensor data and extracted features in noisy environments. 
A Fine Gaussian Support Vector Machine (SVM) 
was chosen for predicting valence and arousal scales due to its memory efficiency.
% as the predictor for the valence and arousal scales for its memory efficiency. 
% The dataset was divided into a 70:15:15 ratio for training, validation, and testing purposes, respectively.
The dataset was split into a 70:15:15 ratio for training, validation, and testing. 
Finally, a fuzzy emotion modeling approach~\cite{mandryk2007fuzzy} mapped the valence and arousal scales onto the valence-arousal space, identifying specific emotions with nuanced representation by considering fuzzy boundaries between emotional states.
% Finally, a fuzzy emotion modeling approach~\cite{mandryk2007fuzzy} was employed to map the valence and arousal scales onto the valence-arousal space, resulting in the identification of specific emotions. 
% This approach allows for a nuanced representation of emotions by considering the fuzzy boundaries between different emotional states.
% We train the SVM using personalized calibrated sensor data features and emotions labels as training samples to perform emotion identification. 

%  We conduct preliminary studies to collect data and create training set. These preliminary studies are explained in detail in evaluation section to follow. The accuracy of classification improves if personalized calibration is used from 96.9\%. Corresponding confusion matrix is as shown in Figure \ref{C}.
%Emotion identification is done in a desktop with sensor data retrieved from the cloud database.

% personalized calibration improves emotion recognition efficiency.
% mitigate emotional variation of users owing to varying baseline and threshold
% Making algorithm generalized for all the user and hence very adaptive. 

\subsection{MDP Control with Collective Emotion}

% As discussed in design, our feedback control system involves making decisions of change of speed or content based on the current emotion state and profile of the student. Our system tries to bring the student to the optimal emotion state.
%Here, we describe how Sensemo sustains students to their optimal emotion states by controlling teaching content and pace according to their emotional feedback. The emotion recognition algorithm detects the emotions of all students, then the results are used in a discrete MDP to make control decisions.
% \subsubsection{Structure and Asymptotics of Markov Chain}
% The structure of Markov process used for decision making for Sensemo is shown in the Figure \ref{MDP structure}. It is very important to know if 
% We note that MDP used for decision making is stable and has unique solution for policy. It shall be noted that the input to MDP is state of class emotion. It is non-trivial considering different characteristics of students. There are multiple solutions to obtain collective emotion using individual student emotions, e.g., weighted average, the mode of all emotions. In our methodology, weighted average approach is used:
The input to the MDP is the collective emotion state from individual student emotions, which is
% In our approach, w
defined as the weighted average of the individual emotions. 
% \begin{equation*}  
% s^*\approx\frac{\sum_{i=1}^{N}W_i s_i}{\sum_{i=1}^{N}n_i}
% \end{equation*}
It is computed as $s^*\approx(\sum_{i=1}^{N}W_i s_i)/(\sum_{i=1}^{N}n_i)$,
where
% $s^*$ is the collective emotional state, 
$n_i$ is the number of students in emotional state $s_i$, and $N$ is total number of emotions.
\begin{figure}[t]
\centering
\includegraphics[width = .8\linewidth]{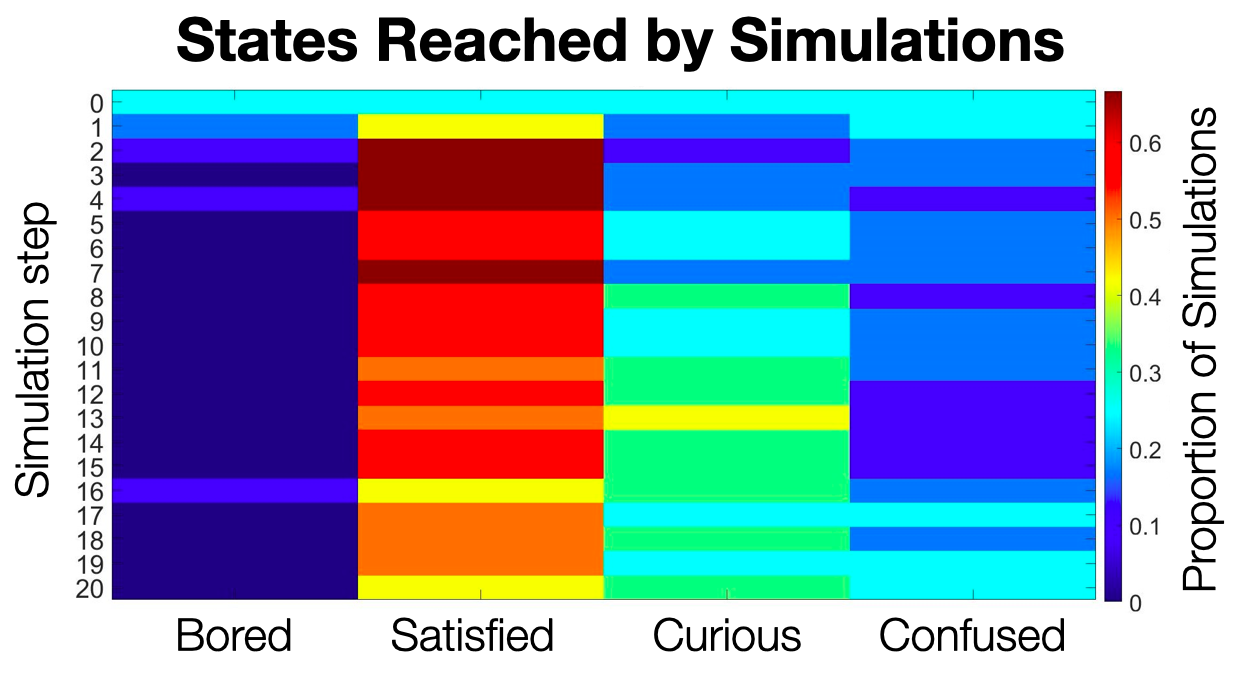}
\vspace{-1ex}
\caption{Convergent asymptotic
behavior of the MDP.}
\label{MDP evol}
\vspace{-0.2in}
\end{figure}
To ensure continuous operation, we analyzed the MDP structure’s stability, irreducibility, aperiodicity, ergodicity, and control process consistency, involving asymptotic behavior and eigenvalue analysis.
% we conducted an analysis of the MDP structure's stability, irreducibility, aperiodicity, ergodicity, and consistency with the control process. 
% This analysis involved examining the asymptotic behavior and performing eigenvalue analysis. 
Figure~\ref{MDP evol} shows a convergence sample of the Markov chain’s asymptotic behavior. 
% illustrates a convergence sample that demonstrates the asymptotic behavior of the Markov chain. 
By analyzing asymptotics with our data, we determined state transition values and computed the stationary state distribution from a uniform distribution.
% By conducting asymptotic analysis and utilizing our experimental data, we determined the state transition values and computed the stationary state distribution, which evolved from a uniform distribution.

% We analyze the asymptotic behavior and eigenvalue analysis to check if MDP structure was stable, irreducible, aperiodic, ergodic and consistent with the control process in order to achieve continuous operation. A convergence sample for asymptotic behavior of Markov chain can be seen in Figure \ref{MDP evol}. Asymptotic analysis was used to find state transition values from our experimental data. We compute stationary state distribution evolved from a uniform distribution. 
% However, it is evident that periodicity of the states satisfied and curious state distribution from settling. Yet it shows allows us to asses range of state transition probability. 

Table~\ref{tb1} shows an optimal and a sub-optimal policy for \sys, helping it adjust decisions based on value iteration results.
% illustrate an example of an optimal policy and a sub-optimal policy for \sys.
% It helps \sys to adjust its decisions based on the values obtained through value iteration.
Note that Table~\ref{tb1} represents one optimal policy, which can vary with data, iterations, reward schemes, and student preferences.
% represents one of the optimal policies, and it can vary depending on factors such as data, number of iterations, reward schemes, and student preferences.
% \sys stores sub-optimal solutions that have rewards lower than the optimal policy, allowing for the generation of alternative suggestions. 
\sys stores sub-optimal solutions with lower rewards than the optimal policy for alternative suggestions.
These are used when optimal suggestions cannot be implemented.
% These sub-optimal policy suggestions are only utilized by \sys when the optimal suggestions cannot be implemented.
For instance, if confusion is the collective emotion, \sys may suggest content simplification.
% if there is a collective emotion of confusion, the \sys system may suggest a simplification in content. 
If infeasible, it proposes sub-optimal adjustments like reducing the teaching pace.
% However, if that suggestion is not feasible, \sys will instead propose sub-optimal policy adjustments, such as decreasing the teaching pace. 
We observe that the sub-optimal policy can change depending on the prevailing conditions, while the optimal policy remains stable.
% The idea here is to use optimal policy as well as 
% Hence, in order to encounter the variability and capability of implementation. Sensemo stores the sub-optimal solutions whose rewards are less than that of optimal policy to generate an alternative.
% solutions cumulative reward lesser rewards than the optimal rewards in order to generate the look up table so that quick adaption of the decision making process.  
% Sub-optimal policy suggestion will only be adopted by Sensmo when optimal suggestion are not implementable. For instance, Sensemo system hints at change of content when there is insufficient content left. Sensemo will instead suggest the sub-optimal policy adjustments of decreasing pace. It shall be noted that sub-optimal policy is subject to change based on conditions while optimal policy is stable.

\subsection{Discussion of System Generalization}
In \sys, personalized calibration enables our approach to be applied to diverse groups of individuals with varying baselines, ensuring its adaptability. 
In addition, the MDP-based decision-making framework can be easily extended to accommodate different numbers of states and objectives, allowing for flexible customization of the state transition probabilities in a generalized system function.
For example, \sys can be utilized for monitoring mental states in medical systems~\cite{shao2019analytical}. 
The methodology employed for emotion recognition in our approach is also capable of incorporating a broader range of emotion states within the valence-arousal space. 
While we utilized a simple four-quadrant classification of emotional space in our case, the number of emotions can be expanded, allowing for further classification of emotional subspaces.
% The system design for Sensemo has mainly consist of personalized calibration, emotion recognition and MDP-based policy making. Personalized calibration allows the methodology to be used on different group of people with varying baselines. Moreover, the MDP-based decision making chain can easily extend to different number of states and objectives of the state transition probabilities of given generalized system function. For instance, Sensemo can be applied in medical field such as mental state monitoring~\cite{shao2019analytical,komorowski2016markov}. The methodology for emotion recognition used in our approach is also able to incorporate more emotion states in valence arousal space. For our case we have used a simple four quadrant classification of emotional space but number of emotions can be varying further classification of subspace. This adaptive features allow the methodology of Sensemo to be easily modelled for various application dealing with emotion recognition and probabilistic decision making.

\begin{table}[t]
\caption{Optimal and sub-optimal policies based on asymptotic value iterations for MDP}
\centering
\label{tb1}
\begin{tabular}{|c|c|c|}
\hline
\textbf{}                                                           & \multicolumn{2}{c|}{\textbf{Policy Based on Value Iteration}}                                                                                                                                                                         \\ \hline
\textbf{Current State $s$} & \textbf{Optimal Policy $\pi(s)$} & \textbf{Sub-optimal Policy $\pi^*(s)$} \\ \hline
bored                                                               & Enriching content                                                                                                        & Simplifying content                                                                                                      \\ \hline
satisfied                                                           & Making no change                                                                                                       & Decreasing pace                                                                                                          \\ \hline
confused                                                            & Simplifying content                                                                                                  & Decreasing pace                                                                                                          \\ \hline
curious                                                             & Decreasing pace                                                                                                      & Enriching content                                                                                                            \\ \hline
\end{tabular}
\vspace{-0.0in}
\end{table}

\section{System Evaluation}
In this section, we evaluate the performance of \sys in emotion recognition and its impact on improving learning outcomes.
To conduct the evaluation, we recruited 22 graduate students between the ages of 22 and 27.
Additionally, all experiments described in this paper received approval/waiver from the IRB.
% All experiments conducted in this paper received an IRB for ethical approval, which is allowed according to the Standard Operating Procedures at authors' affiliated universities.

\subsection{Emotion Recognition Evaluation}

% We evaluate the emotion recognition subsystem on two parameters - accuracy and latency. 
% The emotion feedback data is the bases of our system and thus accuracy of this data and quick response of emotion recognition system is essential.

\noindent \textbf{Experimental Setup}.
To collect data for different emotion states, it is crucial to use stimuli that are consistent and reliable throughout the experiment.
%\sys utilizes the IAPS image dataset, which consists of 1182 pictures of which valence and arousal scales are rated by researchers. 
%These scales serve as suitable target values for mapping and classifying physiological features into the valence-arousal space. 
%Similar approaches have been mentioned in previous studies~\cite{choi2017heart,deak2010hungarian}, where the IAPS was employed for studying emotions and validating the use of heart rate as an effective indicator for emotion identification.
In order to establish a neutral emotional state for volunteers, we utilizes the IAPS image dataset. We present the user with a series of five images that have medium valence and low arousal, lasting for 15 minutes. Following this, the desired images are displayed for data collection purposes. Between subsequent data collection intervals, we again show the user images with medium valence and low arousal. It is important to note that images with low valence and high arousal, which may contain negative content such as mutilation, can have a long-lasting emotional impact and consequently influence subsequent data collection. For this reason, data collection involving such images is performed towards the end of the experiments.

\begin{figure}[t]
    \centering
    \begin{subfigure}[b]{0.48\columnwidth}
        \centering
        \includegraphics[width=\linewidth]{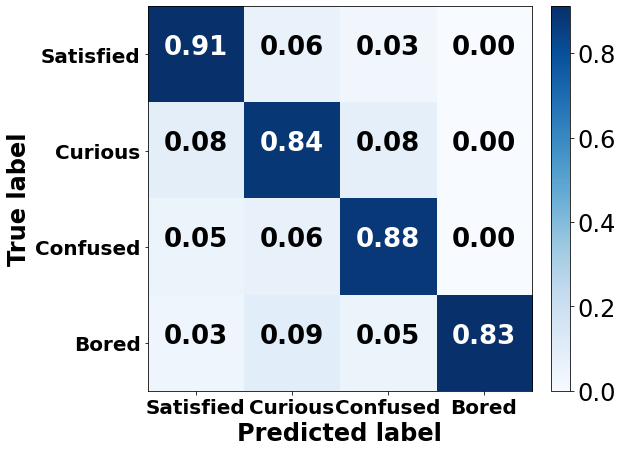}
        \label{fig:confu_mat_2}
            \end{subfigure}\hfill
    \begin{subfigure}[b]{0.48\columnwidth}
        \centering
        \includegraphics[width=\linewidth]{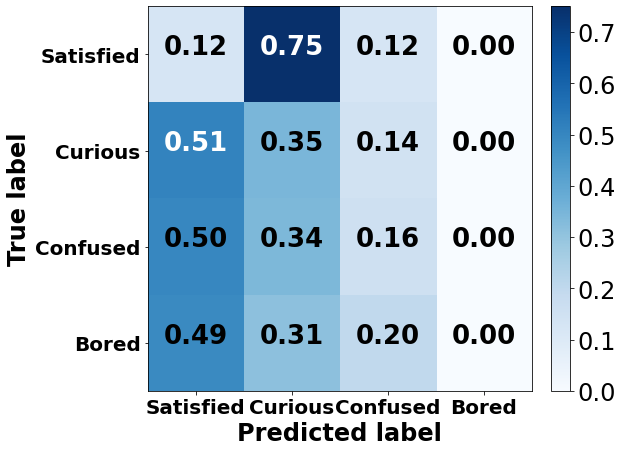}
        \label{fig:confu_mat_1}
            \end{subfigure}
            \vspace{-2ex}
    \caption{Emotion recognition confusion matrices of \sys with (left) and without (right) personalized calibration.}
    \vspace{-2ex}
    \label{fig:confu_mats}
\end{figure}

% \subsubsection{System Accuracy}
% \begin{table}[]
% \caption{Confusion Matrix for Emotion Recognition}
% \centering
% \begin{tabular}{|c|c|c|c|c|}
% \hline
% \multicolumn{5}{|c|}{\textbf{Predicted Class (Uncalibrated)}}                                    \\ \hline
% \textbf{True Class} & \textbf{Bored} & \textbf{Satisfied} & \textbf{Curious} & \textbf{Confused} \\ \hline
% \textbf{Bored}      & 157            & 10                 & 9                &                   \\ \hline
% \textbf{Satisfied}  &                & 138                & 7                &                   \\ \hline
% \textbf{Curious}    &                &                    & 359              &                   \\ \hline
% \textbf{Confused}   &                &                    & 12               & 116               \\ \hline
% \multicolumn{5}{|c|}{\textbf{Predicted Class (Calibrated)}}                                      \\ \hline
% \textbf{Bored}      & 101            & 1                  & 10               &                   \\ \hline
% \textbf{Satisfied}  &                & 134                & 6                &                   \\ \hline
% \textbf{Curious}    &                &                    & 443              &                   \\ \hline
% \textbf{Confused}   &                &                    & 11               & 103               \\ \hline
% \end{tabular}
% \label{C}
% \end{table}
% \vspace{0.05in}
\noindent \textbf{System Accuracy}.
% The accuracy of emotion recognition is important towards the goal of the system. 
% The IAPS photo-set helps in evaluating the accuracy of the emotion recognition subsystem, in addition to forming the training set. 
% The above mentioned experiment setup is used to record data for each picture from the set. 
% To form the learning set, we choose pictures from all the different ranges of valence and arousal ranging from very low to very high. We collect data for 200 pictures from the dataset with 100 pictures overlapping among the volunteers and 100 distinct pictures to cover different range of valence and arousal. 
% All the volunteers were graduate students and followed the above mentioned procedure.
To create the dataset, we carefully select IAPS images that span a wide range of valence and arousal levels, encompassing both very low and very high values.
From the dataset, we gather data for a total of 200 images. Among these, 100 images are shared among the volunteers, while the remaining 100 images are distinct, covering various ranges of valence and arousal. 
This compilation forms our dataset, which includes recorded physiological sensor data along with rated arousal and valence scales.
% This forms the training set with the recorded sensor data and with rated arousal and valence scales. We made sure to balance the number of data points for each range category of arousal and valence- low, mid and high, so that each range had equal number of data points. 
Following personalized calibration, we train a Fine Gaussian SVM with supervised learning.
To evaluate the classification accuracy, we conduct a 10-fold cross-validation. The resulting confusion matrix is shown in in Figure~\ref{fig:confu_mats} (left).
The experimental findings indicate that the Fine Gaussian SVM achieves an overall accuracy of 88.9\%. In comparison, the k-nearest neighbors algorithm (KNN) yields an accuracy of 75.2\%.
% Figure~\ref{fig:confu_mat_2} shows the confusion matrix of the results.
% The experiments show that Fine Gaussian SVM has overall accuracy of 88.9\%. The accuracy of Gaussian SVM is better as compared to k-nearest neighbors algorithm (KNN) that gives accuracy of 75.2\%.  

% \vspace{0.05in}
\noindent \textbf{System Reliability}.
In this evaluation, we examine the reliability of \sys when the users are presented with instantaneous stimuli, i.e., video and music~\cite{silveira2013predicting}.
Specifically, we gather physiological data while users are exposed to video stimuli in the form of music videos that incorporate elements of strong emotional stimuli. It is important to note that these videos are compilations of multiple graphics sourced from the IAPS dataset, which elicit various emotions at different times.
The same set of videos is utilized for all users in order to evaluate if similar patterns of physiological signals emerge across different individuals in response to the same stimuli.
Our findings reveal that the response patterns of GSR and HRV exhibit similarities among different users when experiencing the same emotion.
These results demonstrate that not only can we rely on patterns of physiological signals to infer emotions, but \sys also reliably captures and records such data. 
Furthermore, through personalized calibration, we mitigate differences in the physiological signal responses of different users and enhance the accuracy of emotion recognition.
% Here, we evaluate \sys's reliability with instantaneous stimuli
% % Previous work discusses about the effect of video and music clips on emotion of the users~\cite{hanjalic2005affective,silveira2013predicting}. 
% (e.g., video and music~\cite{hanjalic2005affective,silveira2013predicting}). We collect physiological data when the users are exposed to video stimuli. The videos are music videos that contain certain elements of strong emotional stimuli. It shall be noted that the videos were compilation of multiple graphics from IAPS dataset which will expose user to different emotions at different times. These videos were kept the same for all the users. The aim of this experiment is to find similar pattern of physiological signals from different users for the same stimuli. 
% \begin{figure}[t!]
%     \centering
%     \includegraphics[width=0.8\linewidth]{motive.png}
%     \caption{Physiological response, GSR (top) and HRV (bottom), of users to video stimuli.}
%     \vspace{-0.1in}
%     \label{GSRHRV}
%     \vspace{-0.1in}
% \end{figure}
% We found the response patterns of GSR and HRV to the same emotion are similar among different users. The results show not only we can rely on physiological signal patterns for inferring emotions and also \sys is well capable of recording such data reliably. Moreover, personalized calibration alleviate difference between physiological signal responses of different users and improve emotion recognition accuracy.

% \vspace{0.05in}
\noindent \textbf{Ablation study on personalized calibration}.
Figures~\ref{fig:confu_mats} shows the confusion matrices obtained with and without personalized calibration, respectively. 
Across all emotions, there is a significant and noticeable increase in the true positive rate. 
Additionally, the overall recognition accuracy improves from 27.4\% to 88.9\% with the implementation of personalized calibration.
Throughout our experiments, we observed instances of incorrect and unstable classification of user emotional states when personalized calibration was not applied, even when the user's emotions remained consistent. 
These findings suggest that personalized calibration improves robustness of \sys. 
It is worth noting that similar outcomes were observed for all the selected features. 
These results underscore the importance of personalized calibration in standardizing feature extraction and enhancing emotional recognition.

% Figure~\ref{fig:confu_mat_1} and~\ref{fig:confu_mat_2} are the confusion matrices without and with personalized calibration, respectively. The true positive rate increases drastically for all emotion classes. The overall recognition accuracy greatly increases from 27.4\% to 88.9\% after personalized calibration. In our experiments, we observed incorrect unstable classification of user emotional states without personalized calibration when user emotions are stable. This also suggests that the emotion recognition without personalized calibration is not robust. It shall be noted that similar results were observed for all the features discussed as well as emotional variation, e.g., HR and STR. It shows the importance of personalized calibration in standardizing feature extraction and emotional recognition.

\begin{figure*}[h]
    \centering
    \begin{subfigure}[b]{0.28\textwidth}
        \centering
        \includegraphics[width=\linewidth]{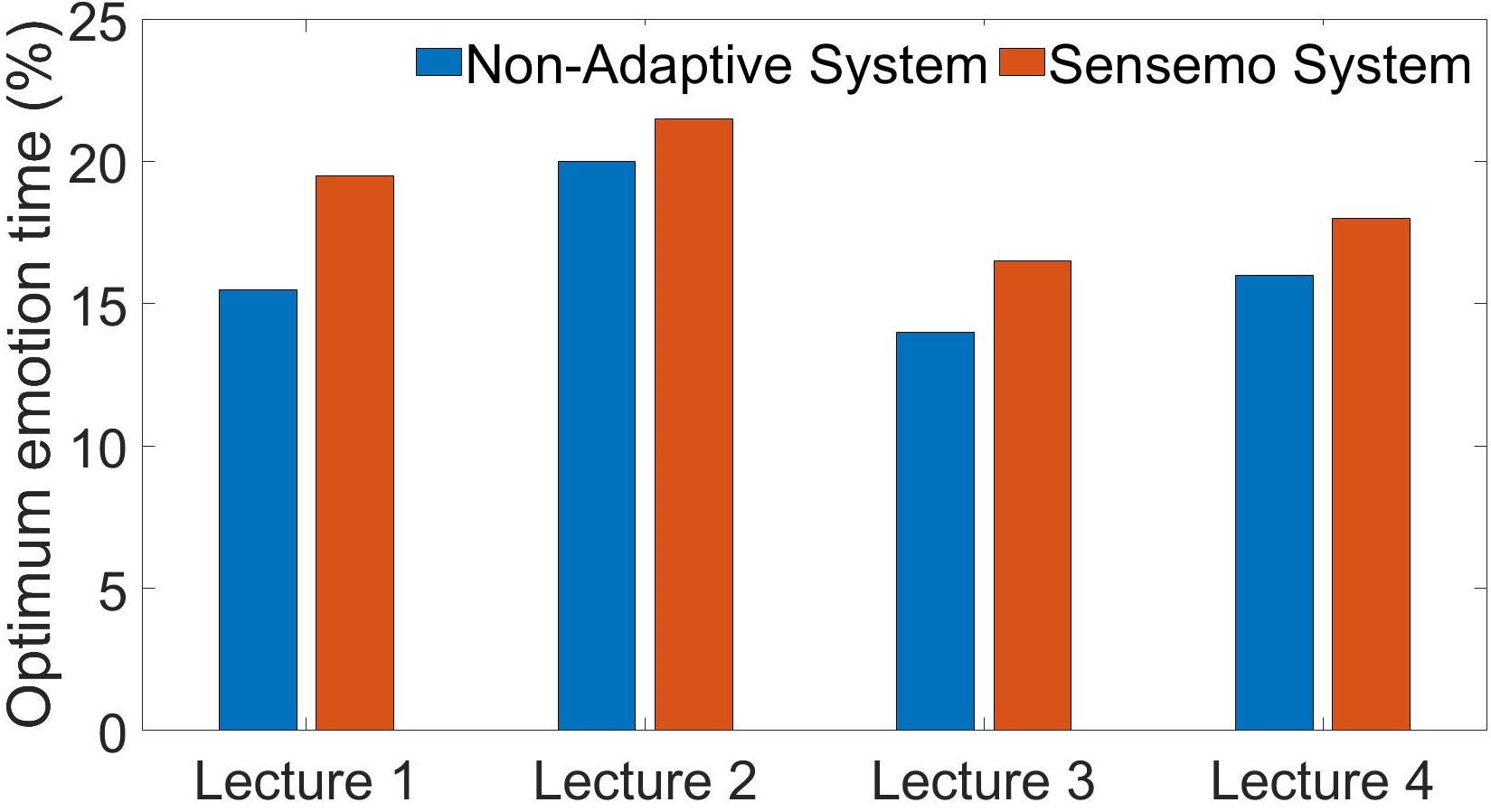}
        \label{SCEMT}
    \end{subfigure}\hspace{0.05\textwidth}
    \begin{subfigure}[b]{0.25\textwidth}
        \centering
        \includegraphics[width=\linewidth]{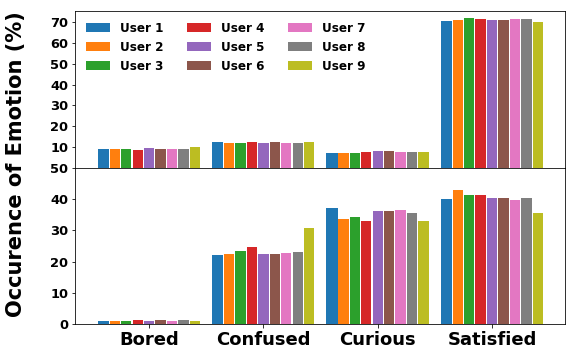}
        \label{fig:response}
    \end{subfigure}\hspace{0.05\textwidth}
    \begin{subfigure}[b]{0.25\textwidth}
        \centering
        \includegraphics[width=\linewidth]{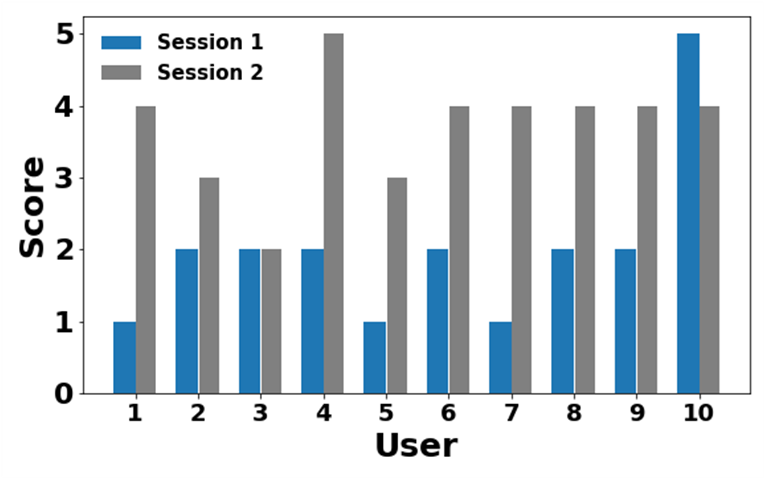}
        \label{QZZ}
    \end{subfigure}
    \vspace{-2ex}
    \caption{Comparison of various aspects in online and in-person classroom settings. (Left) Students using \sys exhibit a longer duration of curiosity emotion in online remote learning setting. (Middle) Emotional responses from all users in two sessions for the classroom evaluation. (Right) Quiz scores of two sessions in in-person classroom setting.}
    \vspace{-3ex}
    \label{fig:comparison}
\end{figure*}

\subsection{Reinforcement Learning-based Controller Evaluation}
The role of the reinforcement learning-based controller is to adjust the system according to the students' current emotional state. It is essential for the controller to respond promptly and efficiently, ensuring it can adapt to the right moment. Consequently, the response time of the feedback control system, and consequently, the entire system, should be minimized.
In addition to response time, another crucial factor is the acceptance of the feedback system by the students. It is important that the students perceive the feedback system positively and find it valuable in their learning experience.
% The reinforcement learning-based controller is responsible for adapting the system based on the current emotion of the students. 
% For the correct and efficient working of the system, it is important that the control provides adaptations for the right moment of emotion detected. In other words, 
% It should be able to respond fast enough to cater to the right moment. Thus, response time of feedback control system and, in effect, of the whole system should be low. Another important parameter is the acceptance of the feedback system by students. 

% \vspace{0.05in}
\noindent \textbf{Response Time}.
We evaluate the response time for the adaptation decision inference using the emotion recognition results.
In this analysis, we also take into consideration the time required by the emotion recognition model. 
By examining the response time, we gain insights into how long it takes for the system to provide a useful suggestion or adaptation once the user's data becomes available. 
To measure the response time, we utilize software timers.
These experiments involve using real lecture slides obtained from the MIT OpenCourseWare~\cite{mit}, where no adaptation is provided to users but only recorded for the purpose of analysis. 
After analyzing the results, we find that the average response time of the feedback system is approximately 3.1 seconds, which is significantly shorter than the duration of any individual lecture slide.
% We test the feedback system for its response time with respect to the inferred emotion data to the adaptation decision. We include the time taken by the emotion classifier into account. Thus, this response time analysis helps understand the time taken by the system to provide a useful suggestion/adaptation once the user's data is made available. We find this response time through software timers. These experiments are conducted by using real lecture slides obtained from the MIT open courseware where no adaptation is provided to users but only recorded for purpose of analysis. The average response time of the feedback system comes out to be 3 sec which is significantly less than the duration of any single lecture slide. 

% \vspace{0.05in}
\noindent \textbf{System Acceptance}.
The purpose of the feedback is to enhance the student's learning experience with minimal effort on their part. To evaluate the effectiveness of the feedback system, we introduce another parameter: the number of manual interventions made by the student to adjust the pace or content.
This parameter serves as a measure of feedback acceptance, where a lower number of interventions indicates better performance of the feedback system. We track these interventions using a software counter.
Our experiments show that, on average, students only make approximately 2.2 interventions per minute. This suggests that the feedback system effectively assists students in maintaining a suitable teaching pace and content, minimizing the need for frequent manual interventions.
% The feedback is meant to facilitate student's learning experience with his/her minimum effort. We create another parameter of evaluation as the number of interventions by the student to adjust the pace or slide of the lecture as test for feedback acceptance. The less the number of interventions are the better the performance of feedback system. We note these interventions through software counter. We find that on average student interventions are 2 per minute, which is acceptable. 

\subsection{Resource Utilization Study}

We evaluate the energy consumption and memory usage of \sys. In our experiments, the estimated battery life for a Microsoft Band 2 with the \sys app connected is 5 hours, which exceeds the duration of a typical class session.
Regarding memory usage, the Android application requires only 3.28 MB of internal memory space. Furthermore, the average runtime memory space utilized by the application is a mere 215 KB. In addition, Table~\ref{tab:tb2} shows a detailed breakdown of power usage while the app runs on Android devices.
% The battery consumption for an Microsoft Band 2 was monitored through native system application, and the estimated battery-life for the watch with the \sys app connected is 5 hours, which is well beyond a typical class session. The android application only takes 3.28 MB of memory space in internal memory. The application, on average, uses just 215 KB of run time memory space, which is really low. Detailed power usage breakdown of app running on Android is as shown in Table~\ref{tab:tb2}.

%The estimated battery-life of an android device, with 2600 mAh battery, running the Sensemo application is 18 hours as the Sensemo app uses 720 mW power on an android device. 

\begin{table}[t]
\begin{center}
\caption{Power usage breakdown of \sys application running on Android devices}\label{tab:tb2}
\begin{tabular}{ |c|c|c|c|c| }
\hline
\bf State & \bf Screen & \bf Bluetooth & \bf Band & \bf Power \\
\hline
Standby & Off & Off & Disconnected & $\leq$ 22 mW \\
\hline
App & Off & On & Disconnected & 273 mW \\
\hline
App & Off & On & Connected & 340 mW \\
\hline
\end{tabular}
\end{center}
\vspace{-0,2in}
\end{table}

\subsection{Real-world Learning Scenarios}\label{subsec:scene}

In this section, we conducted real-world experiments to evaluate \sys’s impact on learning outcomes in both online remote learning and in-person classroom settings.
For the online remote learning, 22 graduate students were randomly divided into two groups of 11. 
One group used \sys while the other did not.
All students wore identical smartwatches and viewed lecture slides from four different lectures (30-50 minutes each) on topics ranging from biology to geography, without any prior knowledge.
Their physiological data was recorded, and comprehension was tested with a quiz of 10 concept questions at the end.
In the in-person classroom setting, 10 graduate students participated in two sessions, wearing identical smartwatches. 
In the first session, the instructor did not receive real-time feedback.
In the second session, the same instructor received real-time feedback from \sys and adjusted the lecture content and pace based on students' emotions. 
Both sessions concluded with a quiz of 5 conceptual questions. 
Students also completed a survey about their emotional states during the sessions.

\noindent \textbf{Learning outcome}.
For the online setting, students using \sys scored 40.0\% higher on the quiz compared to those who did not. Emotion recognition analysis showed longer durations of curiosity among \sys users, suggesting significant potential learning impact, shown in Figure~\ref{fig:comparison} (left).
This suggests that even a modest increase in the time spent experiencing the desired emotion can significantly influence learning outcomes.
Also, \sys's adaptability in teaching pace and content further contributed to improved learning performance.
In the classroom setting, students exhibited more curiosity and less boredom in the session with real-time feedback, shown in Figure~\ref{fig:comparison} (middle). Although confusion increased, it might indicate active learning. Quiz scores, shown in Figure~\ref{fig:comparison} (right), were higher in the second session, hinting at \sys's positive impact despite non-randomization and material control issues. Students also reported that wearing the smartwatch was neither distracting nor inconvenient.

\section{Conclusion}%
%This paper presents ,% a system which can use this personal data to improve user's learning experience tremendously and help improve the learning curve of students by boosting their interest and curiosity into the subject matter. 
In this paper, we present \sys, the first affective learning system that uses real-time data from physiological sensors available in commercial smartwatches.
It achieves an average of 88.9\% accuracy in inferring users' emotions based on valence and arousal scales, within learning environments.
By utilizing sensed emotional states, \sys enables the adaptation of teaching materials and pace. 
The effectiveness of this adaptability has been evaluated in both online remote learning for individual users and in-person classroom settings with multiple users.
Our findings suggest that integrating wearable sensing into affective learning systems has the potential to enhance learning outcomes when compared to traditional approaches.
% \sys is validated in both single user online learning and multi-user classroom settings, assisting students to achieve better learning outcomes over the traditional learning approach. \sys demonstrates good potential to integrate wearable sensing into affective system designs.

% The online learning system is single user system where Sensemo automatically adapts the lecture speed and content based on user's emotions, while also allowing user to manually move back and forth. The classroom system is multi-user based system where the Sensemo provides suggestion to instructor for changing speed or content based on collaborative data of students. We evaluate our system for its ability to keep user in optimal emotional state and through student survey feedback, and find that our system provides improvement over the non-adaptive system by keeping the student in curious state of emotion and is also unobtrusive and non-distracting for students to use based on the survey results. 

\section*{Acknowledgements}

We thank the anonymous reviewers for their valuable feedback and suggestions for improving the quality of this paper.  Research supported in part by NSF CNS-1952096, CNS-1553273 (CAREER), Hong Kong RGC GRF 14201924, and CUHK Direct Grant for Research 4055216.

\bibliographystyle{IEEEtran}
\bibliography{sample-base}

\end{document}